\documentclass{pasj01}
\usepackage{url}
\Received{$\langle$reception date$\rangle$}
\Accepted{$\langle$acception date$\rangle$}
\Published{$\langle$publication date$\rangle$}
\begin{document}

\title{Extended star-forming region within galaxies   
in a dense proto-cluster core at z=2.53\thanks{Based on data collected at Subaru Telescope, which is operated by the National Astronomical Observatory of Japan}}
\author{Tomoko L. Suzuki${}^{1,2}$, Yosuke Minowa${}^{3,4}$, Yusei Koyama${}^{3,4}$, Tadayuki Kodama${}^1$, Masao Hayashi${}^2$, Rhythm Shimakawa${}^3$, Ichi Tanaka${}^3$, Ken-ichi Tadaki${}^2$}%
\altaffiltext{1}{Astronomical Institute, Tohoku University, 6-3 Aramaki, Aoba-ku, Sendai, Miyagi 980-8578, Japan}
\altaffiltext{2}{National Astronomical Observatory of Japan, 
 2-21-1 Osawa, Mitaka, Tokyo 181-8588, Japan}
 \altaffiltext{3}{Subaru Telescope, National Astronomical Observatory of Japan, National Institutes of Natural Sciences, 650 North A'ohoku Place, Hilo, HI 96720, USA}
 \altaffiltext{4}{Department of Astronomical Science, SOKENDAI (The Graduate University for Advanced Studies), 2-21-1 Osawa, Mitaka, Tokyo 181-8588, Japan}
\email{suzuki.tomoko@astr.tohoku.ac.jp}


\maketitle

\begin{abstract}
At $z\sim2$, 
star-formation activity is thought to be high even in high-density environments 
such as galaxy clusters and proto-clusters. 
One of the critical but outstanding issues is 
if structural growth of star-forming galaxies can differ depending on their surrounding environments.  
In order to investigate how galaxies grow their structures and what physical processes are involved in the evolution of galaxies, 
one requires spatially resolved images of not only stellar components but also star-forming regions within galaxies. 
We conducted the Adaptive Optics(AO)-assisted imaging observations for 
star-forming galaxies in a dense proto-cluster core at $z=2.53$ with IRCS and AO188 
mounted on the Subaru Telescope. 
A combination of AO and narrow-band filters allows us to 
obtain resolved maps of H$\alpha$-emitting regions with an angular resolution of 0.1--0.2~arcsec, which corresponds to $\sim1$ kpc at $z\sim2.5$. 
Based on stacking analyses, 
we compare radial profiles of star-forming regions and stellar components and 
find that the star-forming region of a sub-sample with $\rm log(M_*/M_\odot)\sim10-11$ is more 
extended than the stellar component, 
indicating the inside-out growth of the structure. 
This trend is similar to the one for star-forming galaxies in general fields at $z=2-2.5$ obtained with the same observational technique. 
Our results suggest that the structural evolution of star-forming galaxies at $z=2-2.5$ is mainly driven by internal secular processes irrespective of surrounding environments. 
\end{abstract}

\section{Introduction}

High-density environments at $z\gtrsim2$, 
such as galaxy clusters and proto-clusters, 
are often associated with active star-forming galaxies 
(e.g., \cite{venemans07,hayashi12,gobat13,koyama13,kato16,wang16}) 
as opposed to local galaxy clusters, which are dominated by passive elliptical galaxies. 
Environmental dependence of physical quantities of star-forming galaxies at $z\gtrsim2$, 
such as star-forming activity, gas metallicity, and morphology,  
has being investigated in many studies 
(see \citet{overzier16} and reference therein) 
with the aim of understanding 
the early stage of the environmental effects on galaxy formation and evolution.  
However, environmental dependence of structural growth of star-forming galaxies at high redshift has still not been investigated. 
Some studies on the rest-frame optical or rest-frame UV structures of star-forming galaxies 
in (proto-)cluster environments at $z>1$ have already been carried out
\citep{peter07,overzier08,wang16,Allen17,kubo17,shimakawa18,matharu18,socolovsky19}. 
However, it is yet required to directly trace the structures of star-forming regions as well as 
underlying stars (e.g., \cite{nelson16b,tacchella18,belfiore18,ellison18}) 
to investigate how star-forming galaxies build up their structures 
and what physical processes are involved in their evolution.

Structural growth of star-forming galaxies in general fields is investigated 
with the redshift evolution of stellar size (e.g., \cite{trujillo06,franx08,vdwel14}; Shibuya et al. \yearcite{shibuya15}). 
A positive correlation between stellar mass and size as well as 
a slow redshift evolution of size 
indicate that galaxies increase their stellar masses generally 
by adding new stars to the outer regions, i.e., inside-out growth (e.g., \cite{trujillo06}).   
The structural growth is also investigated by directly tracing on-going star formation within galaxies 
(e.g., \cite{nelson16b,tacchella18,belfiore18,ellison18}). 
In \citet{nelson16b}, they showed that star-forming galaxies at $z\sim1$ grow their structures from inside to outside 
by mapping the spatial distribution of H$\alpha$-emitting region.

From theoretical studies, 
a ``compaction'' event is suggested as an evolutionary path of galaxies
especially at higher redshift, where galaxies are more gas-rich 
\citep{zolotov15,tacchella16a}. 
The compaction event can be triggered by gas-rich major mergers (e.g., \cite{mihos96}) 
or violent disk instabilities of gas-rich disk (e.g., \cite{noguchi99}), 
which drives molecular gas in the disk toward the center and 
induces starburst \citep{dekelburkert14}. 
The star-forming region is centrally concentrated in the compaction event, 
and thus, the spatial distribution of the star-forming region is expected to  
become different from that of the inside-out growth.

The relative contribution among these physical processes may depend on the environments where galaxies reside. 
Gas inflow process to make gravitationally unstable disks 
or the frequency of major mergers  
possibly depend on the surrounding environments. 
Galaxies in cluster environments 
reside in more massive dark matter halos than the field counterparts. 
Such cluster galaxies tend to be fed by more intense cold gas inflow from the outside 
until their dark matter halo masses exceed a critical mass where the mode of gas accretion
would change as predicted by cosmological simulations 
(e.g., \cite{keres05,dekel06,keres09}). 
In high-density environments, the frequency of galaxy mergers can be high 
as reported in \citet{lotz13,hine16}. 
Additionally, galaxies in cluster environments 
can be affected by characteristic physical processes  
in cluster regions. 
Ram pressure stripping \citep{gunngott72} would be effective at least in
local galaxy clusters 
and remove the gas associated with the outer part of galaxies. 
Galaxies in local clusters tend to have centrally concentrated star-forming regions 
(e.g., \citet{koopmann04a} and \citet{schaefer18} for local group-class environments). 
When a dominant physical process is different depending on the environments, 
such a difference is expected to be reflected in the internal structures of 
star-forming regions.

%% This paper
Near-infrared (NIR) integral-field-unit (IFU) observations with the Adaptive Optics (AO) system 
are commonly conducted to obtain spatially resolved star-forming regions for galaxies 
at $z\gtrsim1$ (e.g., \cite{law07,law09,forsterschreiber09,swinbank12,forsterschreiber18}). 
AO-assisted high resolution imaging with narrow-band (NB) filters 
also enable us to trace line-emitting star-forming regions within galaxies in a particular redshift slice.  
The major limitation of the AO-assisted observations is a narrow field of view (FoV). 
However, the observational efficiency can be high in the case of dense cluster cores at high redshifts,
as many galaxies can fall within a single FoV and all of them can be spatially resolved at the same time.
Moreover, by using H$\alpha$ emission line as a tracer of star-forming regions, 
one can trace even relatively dusty star-forming regions 
which tend to be missed by the rest-frame UV observations (e.g., \cite{wuyts13,tadaki14}). 

%% ganba 
We have been conducting $K'$-band and NB imaging observations for star-forming galaxies at $z=2-2.5$
with the Infrared Camera and Spectrograph (IRCS; \cite{tokunaga98,kobayashi00}) 
and AO system (AO188; \cite{hayano08}, \yearcite{hayano10}) on the Subaru Telescope (co-PIs: Y. Minowa and Y. Koyama). 
%% Scientific motivation and advantage of this observational technique
This project aims to spatially resolve strong emission line as well as stellar continuum 
for star-forming galaxies at $z=2-2.5$ across environments, 
and finally to reveal the build-up of stellar structures of galaxies and its environmental dependence. 
Targets of this project are H$\alpha$-selected star-forming galaxies at $z=2-2.5$, 
which are originally obtained by NB imaging surveys, 
namely Mahalo-Subaru 
(Mapping HAlpha and Lines of Oxygen with Subaru: \cite{kodama13}) 
and HiZELS (the High-redshift(Z) Emission Line Survey: \cite{best13,sobral13}). 
The targets consist of galaxies in general fields, 
such as COSMOS and UDS \citep{sobral13,tadaki13}, 
and also in proto-cluster fields \citep{hayashi12,koyama13}. 
Because the same NB filters which were used to detect the H$\alpha$ emitters 
are installed on IRCS, 
we can trace the H$\alpha$-emitting region within the H$\alpha$ emitters 
with an angular resolution of 0.1-0.2~arcsec by AO-assisted NB imaging observations. 
This angular resolution scale corresponds to $\sim1$~kpc at $z=2-2.5$. 
%% Advantage of AO+NB imaging 
By targeting the NB-selected galaxies, whose emission line fluxes are already known, 
we can achieve a high observational efficiency. 
As for the field galaxies, 
20 H$\alpha$ emitters in the COSMOS and UDS field were observed. 
The sample covers a stellar mass range of $\rm log(M_*/M_\odot)\sim9-11$. 
Observational results for the H$\alpha$ emitters in the general fields 
are reported in a different paper (\cite{minowa19}, submitted). 
Here we show results for the H$\alpha$ emitters in a proto-cluster environment at $z=2.53$, 
and compare the results in the proto-cluster and general fields. 
%%%

The paper is organized as follows: 
In Section~2, we describe the target proto-cluster to be studied in this paper,
followed by the details of the observations with Subaru/IRCS+AO188. 
In Section~3, we explain our data analyses to obtain spatially resolved 
stellar continuum maps and emission line maps from the images.
In Section~4, we show our results based on the stacking analyses focusing on
any environmental dependence in the spatial extent of star-forming regions within the galaxies at $z=2-2.5$.
We summarize this study in Section~5. 

Throughout this paper, 
we assume the cosmological parameters of $\rm \Omega_m = 0.3$, $\rm \Omega_{\Lambda} = 0.7$, and $H_{\rm 0} = 70 {\rm km\ s^{-1}\ Mpc^{-1}}$. 
All the magnitudes are given in an AB system, 
and we use the Chabrier initial mass function (IMF; \cite{chabrier03}), 
unless otherwise noted.

\section{Target field and Observation
\label{sec:targets}}

\subsection{A dense proto-cluster core at $z=2.53$
\label{subsec:1558}}

Our target field is a proto-cluster, USS1558-003 (hereafter USS1558), 
which is associated with a radio galaxy at $z=2.53$ \citep{kajisawa06}.
NB imaging observations for this proto-cluster were conducted 
with Subaru/Multi-Object InfraRed Camera and Spectrograph (MOIRCS; \cite{ichikawa06}). 
So far, 107 H$\alpha$ emitters are identified in this proto-cluster based on color-color selections
(Hayashi et al.\ \yearcite{hayashi12}, \yearcite{hayashi16}; \cite{shimakawa18}).

\begin{figure}[t]
    \begin{center}
    \includegraphics[width=0.9\columnwidth]{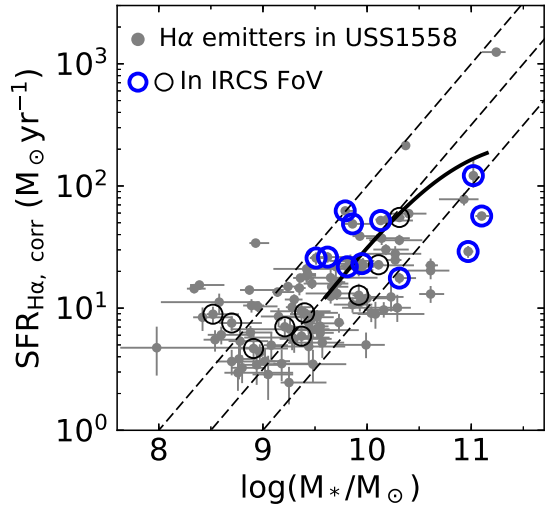}
    \end{center}
    \caption{Stellar mass--SFR relation for the H$\alpha$ emitters in USS1558 
    (Hayashi et al.\ \yearcite{hayashi12}, \yearcite{hayashi16}; \cite{shimakawa18}). 
    Open circles show the H$\alpha$ emitters covered in the IRCS FoV. 
    The blue thick ones correspond to the H$\alpha$ emitters analyzed in this study (subsection~\ref{subsec:obs}). 
    Dotted lines represent constant sSFRs ($\rm SFR/M_*$), namely, $\rm log(sSFR\ [yr^{-1}]) = -8.0, -8.5, and -9.0$. 
    The thick solid line shows the star-forming main sequence at $z=2.53$ from \citet{tomczak16}. 
    Our IRCS+AO188 targets are located around the ``main sequence'' of star-forming galaxies at this epoch. 
    }
    \label{fig1:MS}
\end{figure}

The stellar mass--SFR relation of the H$\alpha$ emitters in USS1558 is shown in figure~\ref{fig1:MS}. 
%% Stellar mass estimate 
Here stellar masses were estimated from SED fitting using 
seven broad-band photometries, namely, $B, r', z', J, H, F160W$ and $K_s$, 
by \citet{shimakawa18}.
%%% 
Because we use a different method to estimate dust extinction for H$\alpha$ ($\rm A_{H\alpha}$)
from previous studies (Hayashi et al.\ \yearcite{hayashi12}, \yearcite{hayashi16}; \cite{shimakawa18}), 
we re-calculate SFRs of the H$\alpha$ emitters. 
SFRs are calculated  
with the relation between SFRs and H$\alpha$ luminosities 
of \citet{kennicutt98b} 
considering the difference between Chabrier and Salpeter IMF \citep{salpeter55}. 
When calculating the H$\alpha$ luminosity of the individual emitters from the MOIRCS NB imaging data, 
we subtract a [{\sc Nii}] flux from a total NB flux 
by estimating a [{\sc Nii}]/H$\alpha$ ratio from an empirical relation 
between [{\sc Nii}]/H$\alpha$ and the equivalent-width (EW) of H$\alpha$+[{\sc Nii}] \citep{sobral12}. 
$\rm A_{H\alpha}$ is estimated 
with an empirical relation among $\rm A_{H\alpha}$, EW of H$\alpha$ in the rest-frame, 
and a ratio between observed SFRs estimated from H$\alpha$ and UV established by \citet{koyama15}. 
More details of this method to estimate $\rm A_{H\alpha}$ is mentioned in subsection~\ref{subsec:dustextinction}. 
The H$\alpha$ emitters in USS1558 show a positive correlation between stellar masses and SFRs, known as the ``main sequence'' of star-forming galaxies 
(e.g., \cite{noeske07,daddi07,kashino13,tomczak16}), 
as already reported in previous studies 
(Hayashi et al.\ \yearcite{hayashi12}, \yearcite{hayashi16}; \cite{shimakawa18}).

The spatial distribution of the H$\alpha$ emitters 
in USS1558 is characterized by several groups of H$\alpha$ emitters.
The densest group is located at 1.5~Mpc away from the radio galaxy (figure~5 in \citet{hayashi12}), 
and its local surface density is 30~times higher than that in the general field \citep{hayashi12}. 
We observed this densest group with Subaru/IRCS+AO188. 
In total, 20 H$\alpha$ emitters are covered in the IRCS FoV ($53\times53\ {\rm arcsec^2}$).

\subsection{Observation with Subaru/IRCS+AO188
\label{subsec:obs}}

\begin{table*}[tb]
\caption{Summary of the obtained data in USS1558 with IRCS+AO188. The limiting magnitude is obtained with a 0.5~arcsec diameter aperture.}
    \centering
    \begin{tabular}{cccccc}\hline
    Band & Dates & Exposure time & FWHM & $m_{\rm lim, 3\sigma}$ & AO mode \\
    & (UT) & (hours)  & (arcsec) & (mag) \\ \hline
    $K'$ & 2013 May 7  & 2.2 & 0.25 & 25.55 & LGS  \\
    NB2315 & 2014 May 17, 18, 22 & 7.4 & 0.17 &  24.09 & LGS \\ \hline 
    \end{tabular}
    \label{tab:obssummary}
\end{table*}

We conducted $K'$-band and NB imaging observations for USS1558 with Subaru/IRCS+AO188 
on May 2013 and May 2014 (S13A-059 and S14A-019; PI: Y. Koyama). 
We used NB2315 filter ($\rm \lambda_c = 2.314\ \mu m$, $\rm FWHM=0.03\ \mu m$)
to catch H$\alpha$ emission line from galaxies at $z=2.53$. 
The observations were conducted using the laser guide star (LGS) mode \citep{minowa12} 
with a tip-tilt guide star of $R = 14.9\ {\rm mag}$. 
On-source exposure time of each frame was 50 and 200~sec 
for $K'$-band and NB2315, respectively. 

%% Reduction 
Data reduction was conducted using the IRCS imaging pipeline software, which was originally developed by \citet{minowa05} 
and updated with python scripts by Y. Minowa.  
The pipeline follows a standard reduction procedure: 
flat-fielding with sky flat frames, 
distortion correction, 
sky subtraction, 
and combining the frames. 
The reduction procedure was repeated twice to make an accurate bright object mask 
for sky flat frames. 
When combining the frames, 
we excluded the frames with an extremely large full-width-half-maximum (FWHM). 
%% criteria ??
The total exposure time of the final $K'$-band and NB2315 images is 
2.2 and 7.4 hours, respectively. 
The 3$\sigma$ limiting magnitude of $K'$-band and NB2315 is  
25.55 and 24.09~mag with a 0.5~arcsec diameter aperture. 
The achieved PSF size is 0.25 and 0.17~arcsec in FWHM 
for the $K'$-band and NB2315 image (table~\ref{tab:obssummary}).

Whereas 20 H$\alpha$ emitters are covered in the IRCS FoV, 
the H$\alpha$ emitters with $\rm {log(M_*/M_\odot) < 9.5}$ are barely detected in both $K'$-band and NB image. 
Considering signal-to-noise ratios (S/N) of the data, 
we decided to use only relatively massive H$\alpha$ emitters with $\rm log(M_*/M_\odot) \ge 9.5$. 
We also exclude three H$\alpha$ emitters neighboring bright objects. 
In the following sections, we analyze the remaining 11 H$\alpha$ emitters.  
The locus of 11 H$\alpha$ emitters on the stellar mass--SFR diagram is shown in figure~\ref{fig1:MS}, 
demonstrating that our targets are around the ``main sequence'' in USS1558. 
We note that 
our IRCS targets with $\rm log(M_*/M_\odot) = 9.5-10$ are at an upper edge of the main sequence. 
This is related to an observational result that the H$\alpha$ emitters in the dense groups in USS1558 
tend to have higher SFRs than those in less dense regions with similar stellar masses 
\citep{shimakawa18}.

\section{Data analysis
\label{sec:analysis}}

\subsection{PSF matching
\label{subsec:psfmatch}}

\begin{figure*}[t]
    \begin{center}
        \includegraphics[width=1.\textwidth]{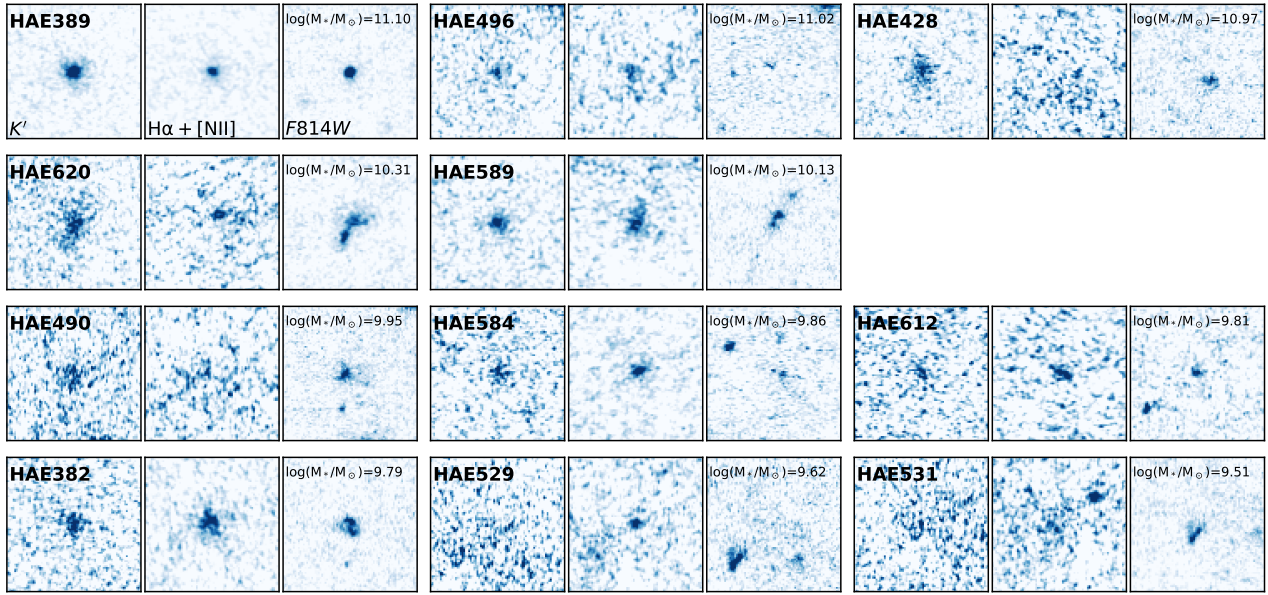}
    \end{center}
    \caption{The IRCS/$K'$, IRCS/NB2315$-K'$, and ACS/$I_{F814W}$ images for 11 H$\alpha$ emitters in the densest group in 
    USS1558 (the image size is $4.3\times4.3\ {\rm arcsec^2}$). 
    The $K'$ and NB2315$-K'$ images correspond to the stellar continuum and H$\alpha$+[{\sc Nii}] emission line, respectively. 
    The $I_{F814W}$ images correspond to the rest-frame 2300 \AA\ at $z\sim2.53$. 
    The IRCS+AO188 images have a similar resolution as {\it HST}. 
    The top two rows show the H$\alpha$ emitters with $\rm 10.0 \le log(M_*/M_\odot)\le11.1$ (high-mass sub-sample).  
    The bottom two rows show the H$\alpha$ emitters with $\rm 9.5 \le log(M_*/M_\odot) < 10.0$ (low-mass sub-sample). 
    IDs are extracted from the catalog by \citet{hayashi16}, 
    and stellar masses are estimated by \citet{shimakawa18}. 
    }
    \label{fig:thumbnail}
\end{figure*}

Because the PSF size of the NB image is smaller than that of the $K'$-band image (table~\ref{tab:obssummary}), 
we smooth the NB image so that we can subtract the $K'$-band image 
from the NB image to obtain continuum-subtracted emission line maps.  
However, comparing the shapes of several stars in the NB image, 
their shapes are slightly elongated in different directions 
probably depending on their positions relative to the LGS. 
We need to conduct the PSF matching for the individual H$\alpha$ emitters according to their position in the FoV.

We fit the $K'$-band and NB images of the seven stars with a moffat function using GALFIT \citep{galfit} to 
make model PSF images. 
Then, 
we create PSF convolution kernels using the model PSF images of the two bands 
and PSFMATCH task in IRAF environment\footnote{\url{http://iraf.noao.edu/}}. 
The NB images of the individual H$\alpha$ emitters are smoothed with the PSFMATCH
by convolving the kernel created from the model PSF images of the closest star.  
Because some of the H$\alpha$ emitters at the center of the image 
have no suitable star,  
we create the model PSF images of $K'$-band and NB for such H$\alpha$ emitters 
by taking an average of the model PSF profiles of the seven stars. 

The PSF shapes of the seven stars in the $K'$-band image are almost independent 
from its position in the FoV, 
and thus, 
the PSF shapes of the convolved NB images are thought to be the same among all the H$\alpha$ emitters. 
After the PSF matching, FWHM of the $K'$-band and NB image is 0.25~arcsec, 
corresponding to 2~kpc at $z=2.53$.

By subtracting the $K'$-band image from the PSF-matched NB image 
for the individual H$\alpha$ emitters, 
we obtain their continuum-subtracted H$\alpha$+[{\sc Nii}] emission line images. 
The $K'$-band image itself corresponds to the underlying stellar continuum 
because NB2315 has little overlap with the $K'$-band in wavelength. 
Figure~\ref{fig:thumbnail} shows the $K'$-band and ${\rm NB2315}-K'$ images taken with Subaru/IRCS+AO188 
and the $I_{F814W}$-band images taken with {\it the Hubble Space Telescope}/the Advanced Camera for Surveys 
({\it HST}/ACS; subsection~\ref{subsec:acs}) 
for 11 H$\alpha$ emitters.

%% Image cutout --> NB - Kp 
%% --> Stacking analysis
\subsection{Stacking analysis
\label{subsec:stacking}}
Unfortunately, S/N of the obtained images 
are insufficient to investigate the resolved structures 
for the individual H$\alpha$ emitters. 
We decide to conduct stacking analyses to achieve higher S/N, 
and to discuss the average properties of their structures. 
We divide our 11 H$\alpha$ emitters into two sub-samples according to their stellar masses, 
namely ``high-mass sub-sample'' including five H$\alpha$ emitters 
with $\rm 10 \le log(M_*/M_\odot) \le 11.1$ and 
``low-mass sub-sample'' including six H$\alpha$ emitters with $\rm 9.5 \le log(M_*/M_\odot) < 10$.

%% local sky subtraction 
%% --> total flux density 
Before stacking, we subtract the local sky for the individual targets 
because the local sky background is not completely subtracted in the pipeline. 
The local sky counts are estimated by taking an average of the counts at $r=30-50$~pix from the center of each target. 
Subsequently, 
we make cumulative radial profiles with a circular aperture, 
and determine total flux densities of the $K'$-band and NB2315 
by taking an average of the cumulative values among radii where 
the profile is flat within 1$\sigma$ error range. 
Total continuum flux densities ($f_c$) is obtained from the $K'$-band images. 
Total fluxes of H$\alpha$+[{\sc Nii}] line 
are estimated with the following equation:

\begin{equation}
    F_{\rm line} = \Delta_{NB} \times (f_{NB} - f_c),
\end{equation}

\noindent
where $f_{NB}$ is a total flux density of NB2315  
and $\Delta_{NB}$ is the FWHM of NB2315 filter (0.03~$\mu \rm m$). 
After normalizing $K'$-band and NB--$K'$ images 
with total $f_c$ and $F_{\rm line}$, 
we conduct the median stacking with IRAF/IMCOMBINE task. 
The central position of each target is determined 
as the flux-weighted center of the $K'$-band image.

%% Normarilzed stacked image ---> Images of observed values 
The obtained stacked images are scaled to the median values of 
the continuum flux densities and the observed H$\alpha$ emission line fluxes 
measured with the MOIRCS images \citep{hayashi16,shimakawa18}.  
%% Halpha + N2 --> Halpha 
The stacked images of emission line maps are contributed from the [{\sc Nii}] emission line 
because FWHM of the NB2315 filter covers both H$\alpha$ emission and [{\sc Nii}] emission lines at $z=2.53$. 
Scaling the stacked NB$-K'$ images to the median H$\alpha$ fluxes 
corresponds to assuming a constant [{\sc Nii}]/H$\alpha$ ratio across a galaxy. 
According to \citet{macuga18}, 
in which a deep X-ray observation with {\it Chandra} was conducted 
($L_X > 2 \times 10^{43}\ {\rm erg\ s^{-1}}$), 
none of our targets are detected in X-ray. 
Our sample includes no X-ray Active Galactic Nuclei (AGNs).
Moreover, half of our targets were observed in previous NIR spectroscopic observations (Shimakawa et al.\ \yearcite{shimakawa15a}, \yearcite{shimakawa15}), 
and the emission line widths in the obtained spectra indicate that they are unlikely host AGNs.
When a galaxy does not host AGN, 
the gradient of [{\sc Nii}]/H$\alpha$ can be regarded as the metallicity gradient. 
Considering that the metallicity gradient of star-forming galaxies at $z\gtrsim1$ 
becomes flatter on average than local ones (e.g., \cite{wuyts16,maiolinomannucci18}),  
an assumption of a constant [{\sc Nii}]/H$\alpha$ ratio for our targets at $z=2.53$ 
seems to be reasonable.

%% Flux loss --> appendix 
We note that the total flux densities obtained from the cumulative profiles 
for the individual targets are not necessarily consistent with those measured in MOIRCS 
images ($K_s$ and NB2315). 
The former total flux densities tend to be lower 
than the later ones. 
Our IRCS+AO188 images seem to lose some fluxes because the image depths 
are insufficient to detect extended components especially for the faint H$\alpha$ emitters. 
We explain more details about this flux loss problem in Appendix~1. 
In the following sections, we do not take the missed fluxes into account, 
and we consider that the missed fluxes do not significantly affect 
our results obtained from the stacking analyses.

\subsection{The rest-frame UV images of the {\it HST}
\label{subsec:acs}}

Imaging observations by 
{\it HST}/ACS for USS1558 
were conducted on 2014 July (GO-13291; PI: M. Hayashi). 
The $I_{F814W}$-band images, which correspond to the rest-frame $\sim$2300\AA\ at $z=2.53$, 
are available for all the IRCS targets (figure~\ref{fig:thumbnail}). 
The 3$\sigma$ limiting magnitude 
is 28.7~mag with a 0.2~arcsec diameter aperture 
after correcting for the Galactic extinction of 0.3~mag (Schlegel et al. \yearcite{schlegel98}). 
We match PSF of the {\it HST} image to that of the IRCS $K'$-band image 
as conducted in subsection~\ref{subsec:psfmatch}. 
Subsequently, the smoothed images of the individual H$\alpha$ emitters are 
stacked in the same manner as for the IRCS images. 
The central position of each target is determined with the $K'$-band image. 
The stacked images are scaled to median flux densities of $I_{F814W}$-band. 
We use the rest-frame UV images 
to estimate $\rm A_{H\alpha}$ as a function of 
a distance from the center (subsection~\ref{subsec:dustextinction}).

\section{Results and Discussion
\label{sec:result}}

\subsection{Radially dependent dust extinction of H$\alpha$
\label{subsec:dustextinction}}

We estimate $\rm A_{H\alpha}$ as a function of radius
using the stacked H$\alpha$ images and the stacked $I_{F814W}$-band images (subsection~\ref{subsec:acs}). 
\citet{koyama15} established an empirical relation using local star-forming galaxies from SDSS to estimate $\rm A_{H\alpha}$:

\begin{eqnarray}
{\rm A_{H\alpha}} & = & (0.101 \times {\rm log\ EW_{H\alpha}} + 0.872) \times {\rm log(H\alpha/UV)} \nonumber \\
                  & \ & + \ (-0.776 \times {\rm log\ EW_{H\alpha} + 1.688)}, 
\label{eq:Aha}
\end{eqnarray}

\noindent 
where $\rm log\ EW_{H\alpha}$ is EW of H$\alpha$ emission line in the rest-frame, 
$\rm log(H\alpha/UV)$ is a ratio between the observed SFR obtained from H$\alpha$ and that from
UV with Kroupa IMF \citep{kroupa02} (eq.~(9)--(11) in \citet{koyama15}).

\begin{figure}
    \begin{center}
    \includegraphics[width=0.8\columnwidth]{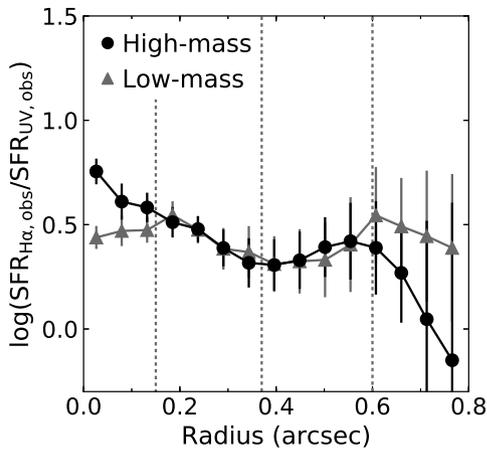}
     \end{center}
    \caption{Radial profiles of the ratio between $\rm SFR_{H\alpha, obs}$ and $\rm SFR_{UV, obs}$ 
    for the two sub-samples.
    At the central region of the high-mass sub-sample, $\rm SFR_{H\alpha, obs}/SFR_{UV, obs}$ becomes higher 
    compared to that of the outer parts.  
    As for the low-mass sub-sample, the SFR ratio is almost flat.
    The vertical dotted lines correspond to the radii, where we take an average of the SFR ratios. 
    The SFR ratios have large uncertainties at $r > 0.6$~arcsec. 
    } 
    \label{fig:SFRobs_ratio}
\end{figure}

We divide the radial profiles of the H$\alpha$ flux by those of the continuum flux density 
to obtain the radial profiles of $\rm EW_{H\alpha}$. 
The radial profiles of $\rm SFR_{H\alpha}$ are then obtained 
from those of $F_{\rm H\alpha, obs}$ with the \citet{kennicutt98b} relation. 
As for the profiles of $\rm SFR_{UV, obs}$, 
we converted the $I_{F814W}$ flux densities ($\sim2300$\AA\ in the rest-frame) 
to SFRs with the \citet{kennicutt98b} relation,  
although \citet{koyama15} used far-UV data rather than near-UV data. 
Figure~\ref{fig:SFRobs_ratio} shows radial profiles of the ratio between $\rm SFR_{H\alpha, obs}$ and $\rm SFR_{UV, obs}$ 
for the two sub-samples. 
The observed SFR ratio seems to increase toward the central region for the high-mass sub-sample, 
whereas the low-mass sub-sample shows almost a flat profile.

When estimating $\rm A_{H\alpha}$ with eq.~(\ref{eq:Aha}),  
we divide the radial profile into three bins, namely, 
$r \le 0.15~{\rm arcsec}$, 
$0.15 < r < 0.37~{\rm arcsec}$, 
and $0.37 \le r \le 0.6~{\rm arcsec}$, 
to achieve higher S/N. 
Because the observed SFR ratios and $\rm EW_{H\alpha}$ have large uncertainties at $r > 0.6~{\rm arcsec}$,
we assume that $\rm A_{H\alpha}$ is constant at $r>0.6$~arcsec, 
and use $\rm A_{H\alpha}$ at $0.37 \le r \le 0.6$. 
The obtained $\rm A_{H\alpha}$ in each radial bin are as follows: 
$\rm A_{H\alpha}$ = 1.11, 0.77, and 0.67~mag for the high-mass sub-sample; 
and $\rm A_{H\alpha}$ = 0.51, 0.46, and 0.40~mag for the low-mass sub-sample. 
Uncertainties on $\rm A_{H\alpha}$ are $0.05-0.1$~mag. 
As expected from figure~\ref{fig:SFRobs_ratio}, 
the central region of the high-mass sub-sample has stronger dust extinction, 
and the low-mass sub-sample has a nearly constant $\rm A_{H\alpha}$. 
The obtained trend is consistent with results of previous studies, 
such as \citet{nelson16a} and \citet{tacchella18}, 
although their results show higher $\rm A_{H\alpha}$, 
for example, $\rm A_{H\alpha}\gtrsim2$~mag at $r<1$~kpc.

We confirm that how to divide the radial profiles with radii 
has little impact on the final dust-extinction-corrected radial profiles in subsection~\ref{subsec:image}. 
Furthermore, as a test, we estimate total $\rm A_{H\alpha}$ for the two sub-samples 
using the median values of $F_{\rm H\alpha}$, $f_c$, and $\rm SFR_{UV,obs}$. 
The total $\rm A_{H\alpha}$ is estimated to be 
0.65~mag and 0.40~mag for the high-mass and low-mass sub-sample, respectively. 
The total values are close to those at $0.37 \le r \le 0.6~{\rm arcsec}$, 
indicating that our $\rm A_{H\alpha}$ measurement for the spatially resolved maps is reasonable.

We note that eq.~(\ref{eq:Aha}) is calibrated with the local star-forming galaxies at $\rm {EW_{H\alpha} < 100}$~\AA, 
whereas the low-mass sub-sample has a median EW of $\sim160$~\AA. 
We extrapolate eq.~(\ref{eq:Aha}) toward a slightly higher EW for the low-mass sub-sample. 
Additionally, 
it is unclear whether the locally calibrated equation is applicable for star-forming galaxies at $z=2.53$, 
because a typical $\rm EW_{H\alpha}$ at a fixed stellar mass changes with redshift \citep{sobral14}.
Although we must keep these facts in mind, 
the methods to estimate the radial dependence of $\rm A_{H\alpha}$ with the currently available data are quite limited. 
Ideally, we would need spatially resolved Balmer decrement ($\rm H\alpha/H\beta$ ratio) maps to estimate $\rm A_{H\alpha}$
more accurately as conducted in \citet{nelson16a}.

\subsection{Stacked stellar mass/SFR maps and radial profiles
\label{subsec:image}}

\begin{figure*}
\begin{minipage}[cbt]{0.24\textwidth}
 \centering\includegraphics[width=1\columnwidth]{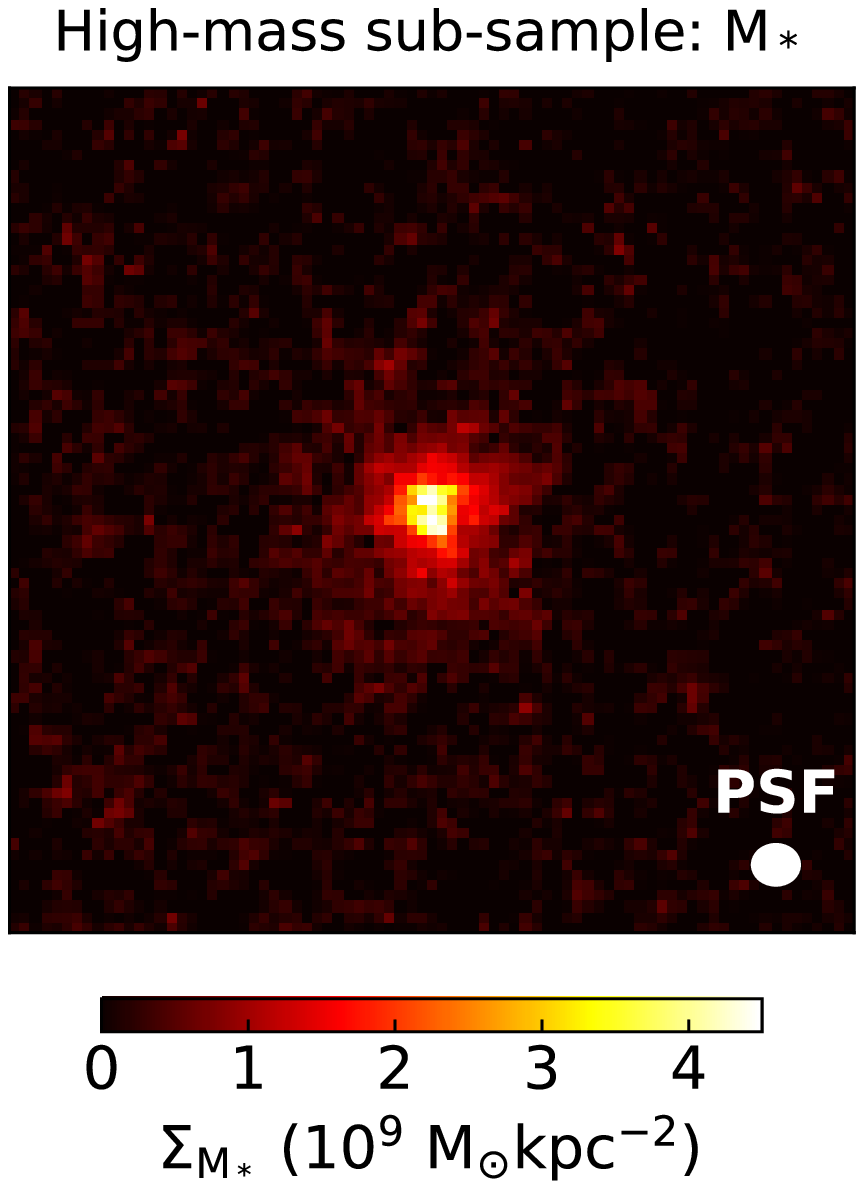}
\end{minipage}
 \begin{minipage}[cbt]{0.24\textwidth}
 \centering\includegraphics[width=1\columnwidth]{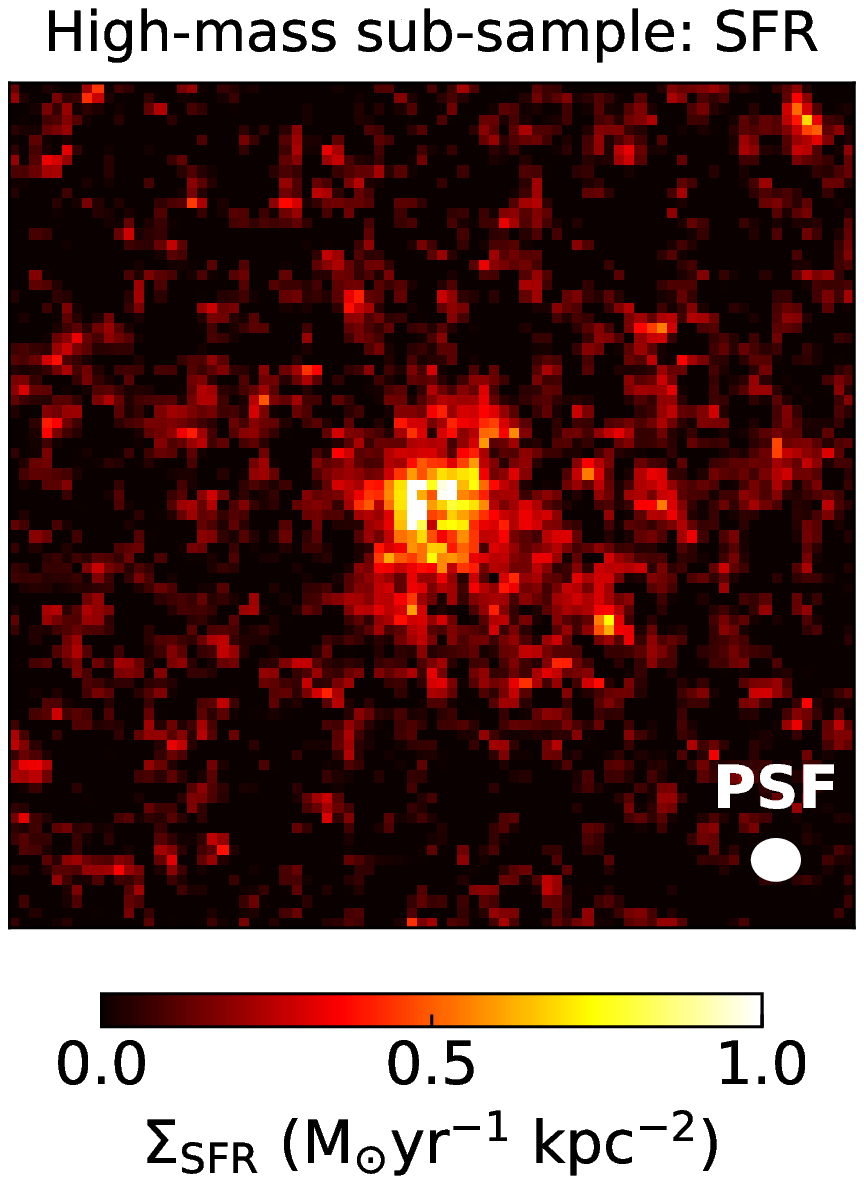}
\end{minipage}
 \begin{minipage}[cbt]{0.24\textwidth}
 \centering\includegraphics[width=1\columnwidth]{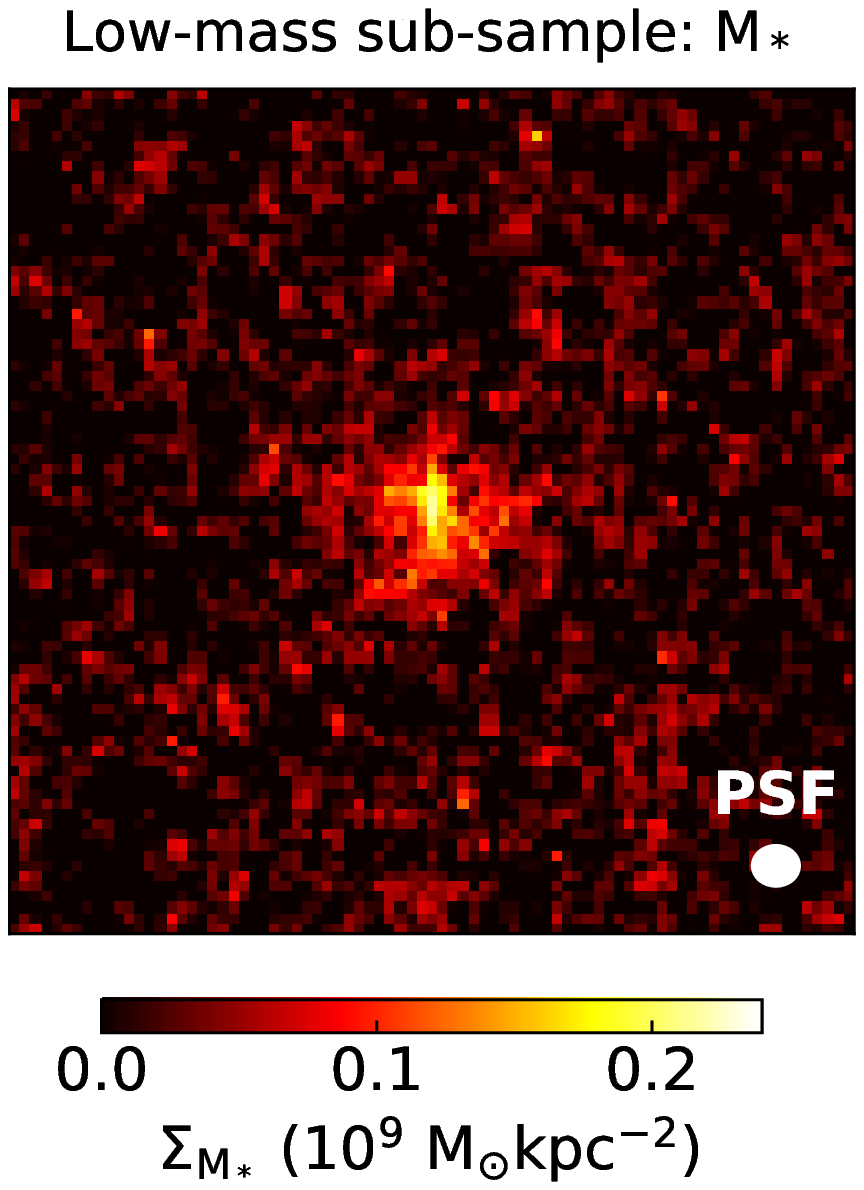}
\end{minipage}
 \begin{minipage}[cbt]{0.24\textwidth}
 \centering\includegraphics[width=1\columnwidth]{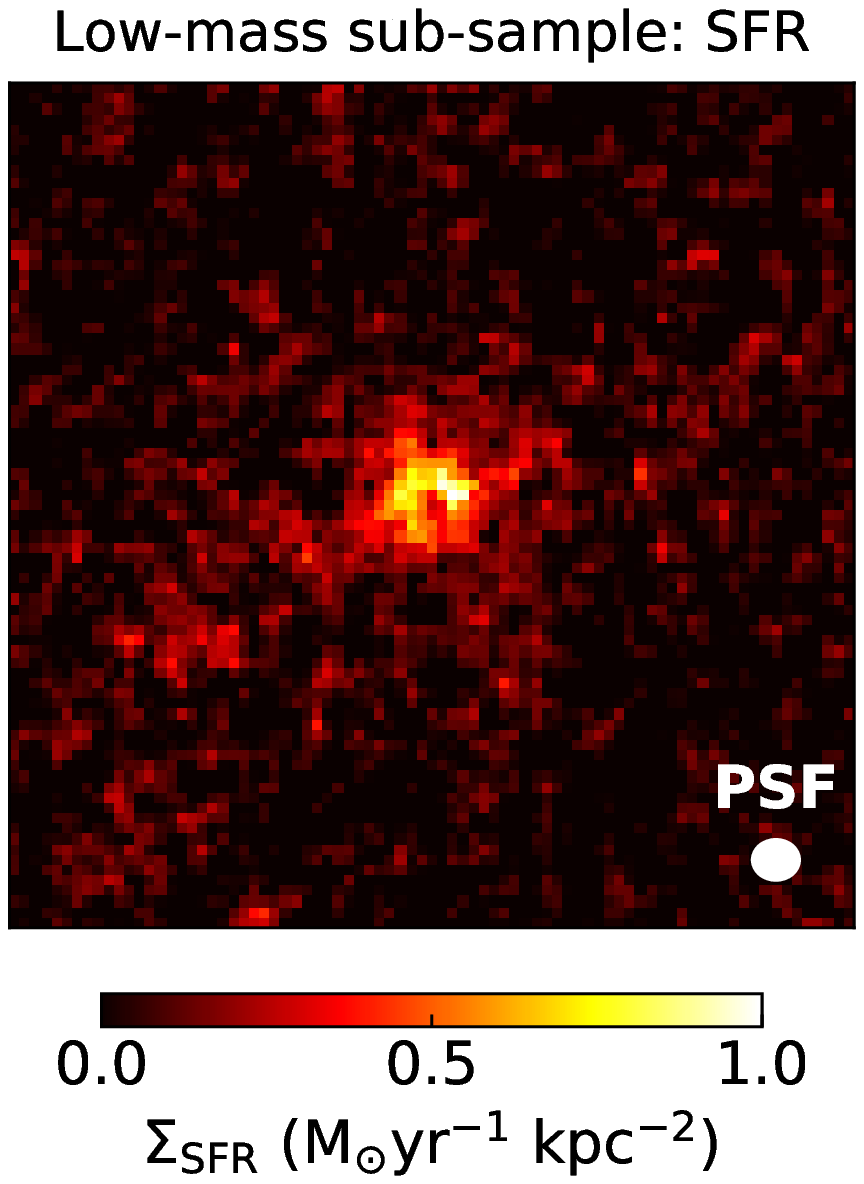}
\end{minipage}
\caption{Stacked images of the stellar mass and $\rm SFR_{H\alpha, corr}$ for the two sub-samples with the radially dependent $\rm A_{H\alpha}$ (subsection~\ref{subsec:dustextinction}). 
The box size of each panel is $\sim 4.3\times4.3~{\rm arcsec^2}$. 
The small white circle in each panel shows the PSF size (FWHM). 
}
\label{fig:stackedimage}
\end{figure*}

\begin{figure*}
    \begin{minipage}[cbt]{0.32\textwidth}
    \centering\includegraphics[width=0.97\columnwidth]{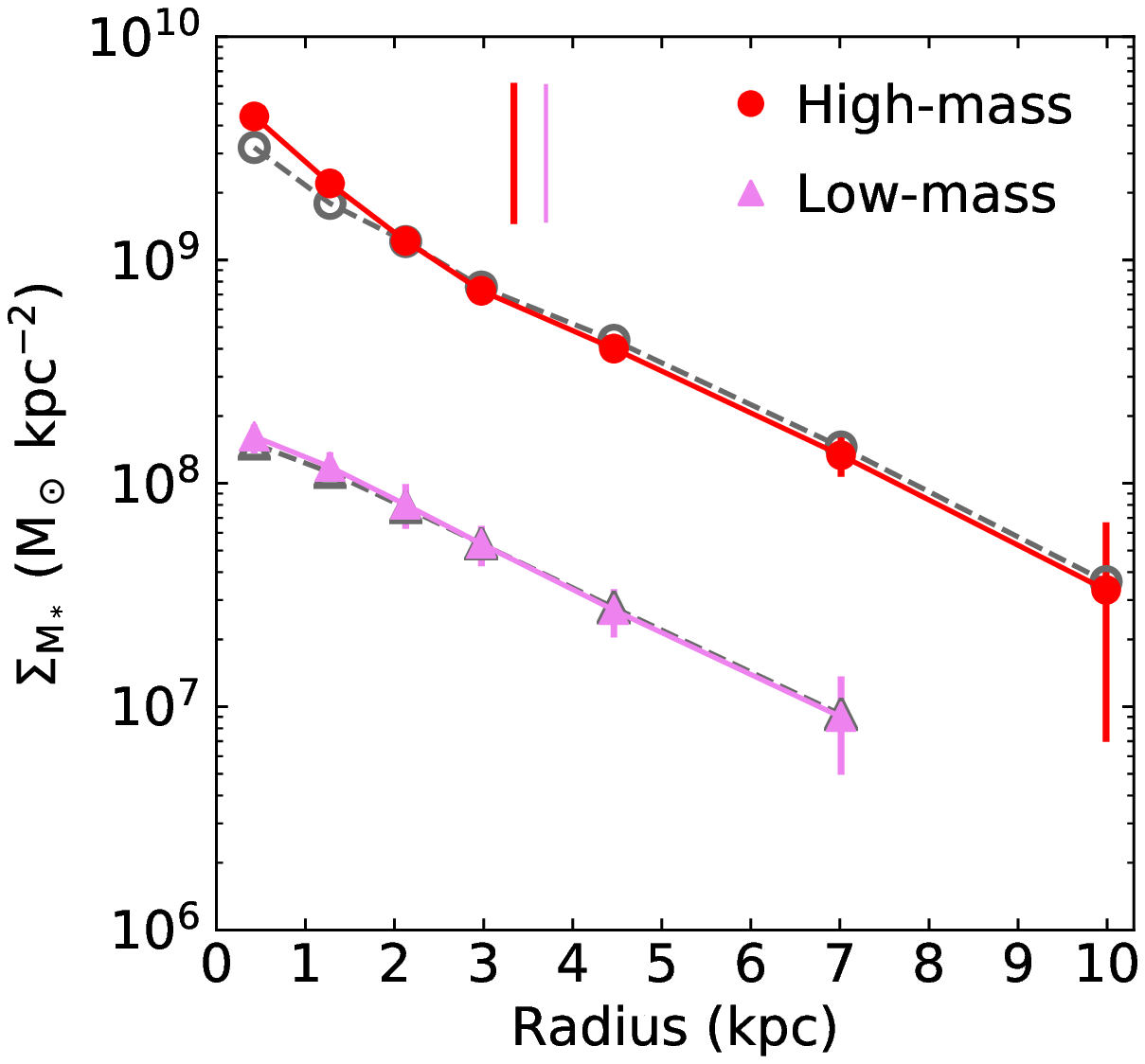}
    \end{minipage}
    \begin{minipage}[cbt]{0.32\textwidth}
    \centering\includegraphics[width=1.\columnwidth]{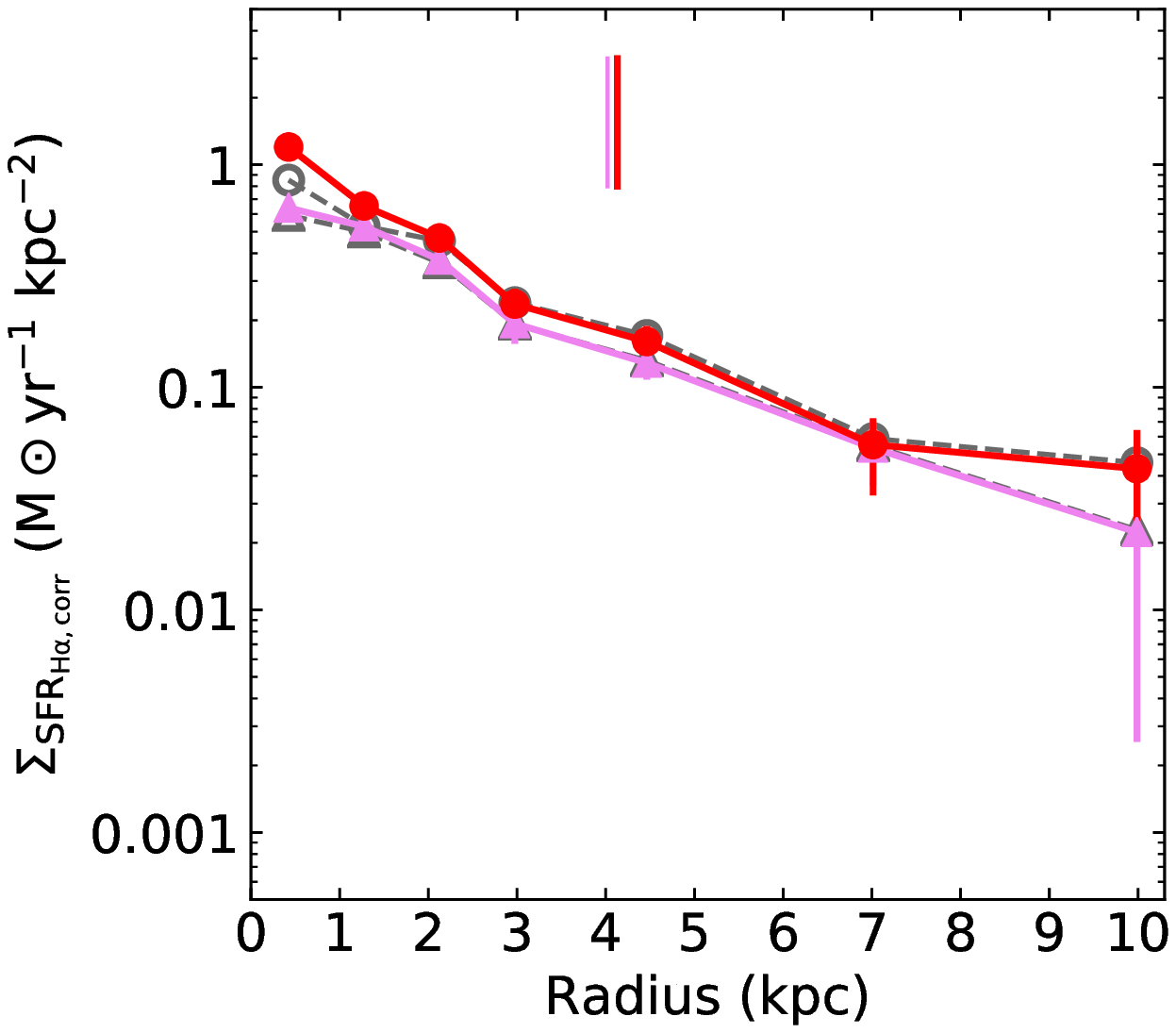}
    \end{minipage}
    \begin{minipage}[cbt]{0.32\textwidth}
    \centering\includegraphics[width=0.95\columnwidth]{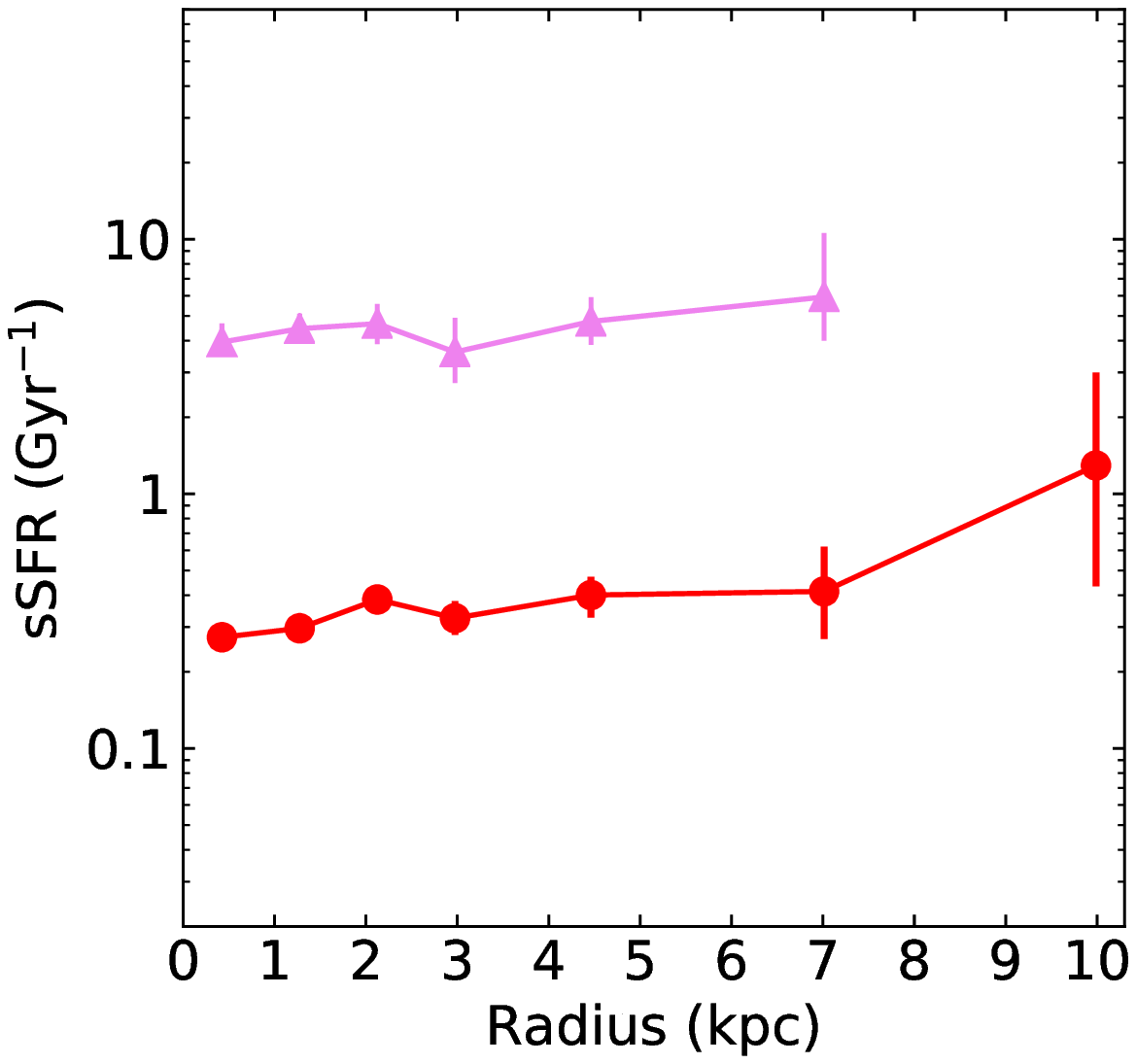}
    \end{minipage}
    \caption{Radial profiles of the two sub-samples.
    From the left to right, each panel shows the radial profiles of stellar mass surface density, those of SFR surface density, and those of sSFR. 
    In the left and the middle panels, 
    the filled symbols represent the results using the radially dependent $\rm A_{H\alpha}$ (subsection~\ref{subsec:dustextinction}). 
    The open symbols represent the results with the uniform $\rm A_{H\alpha}$. 
    The vertical lines in the left and middle panel represent $R_{1/2}$ for the stellar component and star-forming region  
    with the radially dependent $\rm A_{H\alpha}$.
    The thick ones show $R_{1/2}$ for the high-mass sub-sample.
    The SFR profiles of the two sub-samples have a flatter slope than the stellar mass profiles. 
    }
    \label{fig:radialprofile}
\end{figure*}

Figure~\ref{fig:stackedimage} shows the stacked images of the stellar mass and dust-extinction-corrected $\rm SFR_{H\alpha}$ 
for the two sub-samples.
The stacked images of $f_c$ (subsection~\ref{subsec:stacking}) 
are corrected for the dust extinction using radially dependent $\rm A_{H\alpha}$ 
(subsection~\ref{subsec:dustextinction}). 
Here we assume no extra extinction for H$\alpha$ emission line 
compared to the extinction for stellar components, i.e., $f=1$ 
(e.g., \cite{erb06-647}; Reddy et al.\ \yearcite{reddy10}, \yearcite{reddy15}). 
Subsequently, the image of the intrinsic $f_c$ for each sub-sample 
is converted to the stellar mass image 
by multiplying a constant ${\rm M_*}/f_{c,{\rm int}}$ ratio, 
where $f_{c,{\rm int}}$ is the total continuum flux density 
measured from the cumulative profile of the dust-extinction-corrected $f_c$ image. 
Multiplying a constant ${\rm M_*}/f_{c,{\rm int}}$ corresponds to 
assuming a constant intrinsic mass-to-light ratio (${\rm M_*}/L_{\rm int}$) 
across a galaxy. 
Because $M/L$ depends not only on dust extinction but also on age,  
some of our targets can have radial gradients of $M_*/f_c$ even 
after the radially dependent dust extinction correction is applied. 
As for the dust-extinction-corrected $\rm SFR_{H\alpha}$ images, 
we convert the stacked images of $F_{\rm H\alpha}$ obtained in subsection~\ref{subsec:stacking} 
to those of $\rm SFR_{H\alpha}$ with radially dependent $\rm A_{H\alpha}$ and 
the \citet{kennicutt98b} relation. 
We then determine total SFRs for the two sub-samples 
from cumulative profiles of the $\rm SFR_{H\alpha, corr}$ images. 

%% constant A_Ha
For a fair comparison with the field galaxies in subsection~\ref{subsec:comparesize}, 
we also create the images of stellar mass and $\rm SFR_{H\alpha, corr}$ 
with radially constant $\rm A_{H\alpha}$. 
In this case, 
we just scale the observed $f_c$ and $F_{\rm H\alpha}$ images 
so that the total stellar mass and SFR are matched to those derived from the images with radially dependent $\rm A_{H\alpha}$.

Figure~\ref{fig:radialprofile} shows radial profiles of the stellar mass surface density, 
SFR surface density, and sSFR for the two sub-samples. 
Radial profiles of the stellar mass and SFR surface density are obtained by taking an average in an azimuth direction 
with a circular aperture. 
The sSFR profile is obtained by dividing the SFR surface density profile by the stellar mass surface density profile. 
Two kinds of radial profiles are shown for each sub-sample in the left and the middle panels of figure~\ref{fig:radialprofile}, 
which represent the two results using different dust extinction corrections.  
Comparing the radial profiles of the stellar surface density and SFR surface density, 
we find that the SFR profile becomes flatter than the stellar mass profile 
at $r \gtrsim 2-3$~kpc for the two sub-samples.

\subsection{Size comparison between star-forming regions and stellar components
\label{subsec:comparesize}}

Table~\ref{tab:sizesummary} summarizes sizes ($R_{1/2}$) of the stellar mass and $\rm SFR_{H\alpha, corr}$ 
distributions for the two sub-samples. 
$R_{1/2}$ is defined as a radius where a cumulative value becomes a half of the total $\rm M_*$ or $\rm SFR_{H\alpha, corr}$. 
Error bars on $R_{1/2}$ represent 1$\sigma$ scatter of the cumulative values from the total value  
at the outer regions where a change of the cumulative value is 
comparable to or less than a noise level.

We show the two $R_{1/2}$ of $\rm M_*$ and $\rm SFR_{H\alpha, corr}$ 
with different assumptions for dust extinction correction. 
The $R_{1/2}$ with the radially dependent $\rm A_{H\alpha}$ becomes smaller 
than that with the uniform $\rm A_{H\alpha}$ for the high-mass sub-sample 
as expected from the strong dust extinction at the center in the former case (figure~\ref{fig:radialprofile}). 
Regardless of the dust extinction correction methods,
the size measured in the SFR map is larger than 
that measured in the stellar mass maps 
for the high-mass sub-sample, 
which indicates that the star-forming region is more extended than the underlying stellar component. 
Such a further extended star-forming region can be also seen in the comparison of the two radial profiles 
as mentioned in subsection~\ref{subsec:image}. 
As for the low-mass sub-sample, 
$R_{1/2}$ of the two components are the same within 1$\sigma$ error bars
(table~\ref{tab:sizesummary}).

We compare the sizes  
measured with the method mentioned above and IRCS $K'$-band images    
and those measured with GALFIT and {\it HST}/WFC3 $H_{F160W}$-band images \citep{shimakawa18}. 
Here we use the stacked images and the individual images of the bright targets. 
As a result, 
we find that our measurement shows systematically larger sizes by a factor of $\sim1.8$ 
as compared to the measurement with GALFIT. 
Such a systematic difference may be due to the fact that 
our measurement does not deconvolve a PSF and not use a S\'ersic profile to fit the images.  
We note, however, that the relative comparison remains valid as long as we use sizes measured with the same method.

\begin{table*}
\caption{Summary of the total stellar mass, dust-extinction-corrected total SFRs, 
and sizes  
of the stellar components and star-forming regions for the two sub-samples in USS1558. 
}
    \begin{center}
    \begin{tabular}{c|cc|cc|cc} \hline
     & \multicolumn{2}{c|}{Total}  & \multicolumn{2}{c|}{Radially dependent $\rm A_{H\alpha}$} & \multicolumn{2}{c}{Uniform $\rm A_{H\alpha}$} \\ \cline{2-7}
    & Stellar mass & $\rm SFR_{H\alpha, corr}$ & $R_{1/2}({\rm M_*})$ & $R_{1/2}({\rm SFR_{H\alpha, corr}})$ & $R_{1/2}(\rm M_*)$ & $R_{1/2}({\rm SFR_{H\alpha, corr}})$ \\ 
     & $\rm log(M_*/M_\odot)$ & $\rm (M_\odot yr^{-1})$ & (kpc) & (kpc) & (kpc) & (kpc) \\ \hline
    High-mass & 10.97 & 37.77 &  $3.34^{+0.16}_{-0.15}$ & $4.13 \pm 0.11$ & $3.65 \pm 0.16$ & $4.37\pm 0.13$\\
    Low-mass  & 9.80  & 28.23 &  $3.70^{+0.45}_{-0.39}$ & $4.02 \pm 0.23$ & $3.77^{+0.46}_{-0.39}$ & $4.12^{+0.24}_{-0.23}$ \\ \hline
    \end{tabular}
    \end{center}
    \label{tab:sizesummary}
    \begin{tabnote}
     \end{tabnote}
\end{table*}

\subsection{Environmental dependence of mass--size relation
\label{subsec:environment}}

\begin{figure}
    \begin{center}
    \includegraphics[width=0.85\columnwidth]{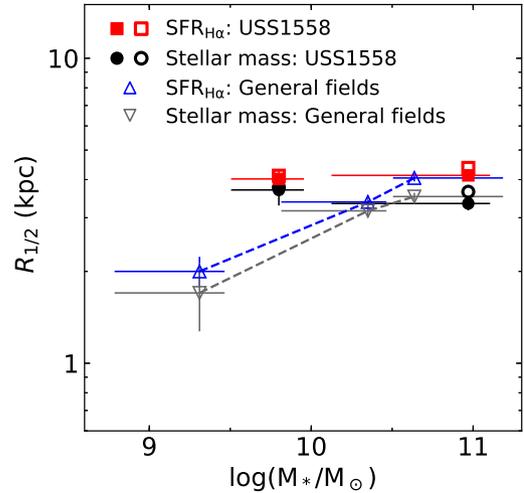}
    \end{center}
    \caption{Relation between the total stellar mass and size of the stellar components and star-forming regions 
    for the H$\alpha$ emitters in different density fields at $z=2-2.5$.
    The sizes measured in the stacked images are shown for the two samples.
    The filled symbols represent $R_{1/2}$ assuming the radially dependent $\rm A_{H\alpha}$, 
    whereas the open symbols represent $R_{1/2}$ obtained with the uniform $\rm A_{H\alpha}$.  
    Both in the general fields and in the proto-cluster core, 
    star-forming regions are more extended than the underlying stellar components at least 
    for the massive star-forming galaxies with $\rm log(M_*/M_\odot) \sim 10-11$. 
    }
    \label{fig:mass_size}
\end{figure}

In Figure~\ref{fig:mass_size}, we compare $R_{1/2}$ measured in the stacked images 
between our sample and the sample of the field galaxies at $z=2-2.5$ 
in \citet{minowa19}. 
%shows the relation between the total stellar mass and $R_{1/2}$ 
%of the stellar components and star-forming regions 
%for the two sub-samples in USS1558 
%and for the field galaxies at $z=2-2.5$ 
%by \citet{minowa19}. 
Their sample consists of 20 H$\alpha$ emitters at $z=2.18, 2.23,$ and $2.53$ 
in the UDS and COSMOS field \citep{tadaki13,sobral13}, 
which cover a stellar mass range of $\rm log(M_*/M_\odot) \sim 9-11$. 
The field sample distributes around the star-forming main sequence at the epoch. 
The same methods are used for stacking and also measuring $R_{1/2}$. 
One difference is that uniform $\rm A_{H\alpha}$ is assumed for the field galaxies.  
In figure~\ref{fig:mass_size}, 
we show the two $R_{1/2}$ with different assumptions in $\rm A_{H\alpha}$ for USS1558 sub-samples. 
We should look at $R_{1/2}$ with the same $\rm A_{H\alpha}$ 
recipe for a fair comparison with the results of the field galaxies.

At least at $\rm log(M_*/M_\odot) \ge 10$, 
the star-forming galaxies in the different environments at $z=2-2.5$ 
have similar sizes for each component 
and show the same trend 
that the star-forming region is more extended than the underlying stellar component 
(\citet{nelson16b} for $z\sim1$ star-forming galaxies in general fields).  
Massive star-forming galaxies at $z=2-2.5$ seem to 
build up their structures from inside to outside 
in spite of their surrounding environments.

As for the low-mass sub-sample, 
the sizes of the stellar mass and SFR are marginally larger than 
those for the field counterparts. 
We find no clear dependence of $R_{1/2}$ on stellar mass for the two sub-samples in USS1558 in contrast to the field sub-samples. 
\citet{shimakawa18} also reported little correlation between stellar mass 
and size for the H$\alpha$ emitters in USS1558 using the {\it HST}/$H_{F160W}$-band images and GALFIT. 
A weak dependence of $R_{1/2}$ on stellar mass in our IRCS targets may be partly due to  
systematically higher sSFRs for the low-mass sub-sample (figure~\ref{fig1:MS}, subsection~\ref{subsec:1558}). 
Indeed, some studies reported a weak positive correlation between sSFRs (or the deviation from the star-forming main sequence) 
and sizes of star-forming galaxies (e.g., \cite{wuyts11_SFMS,whitaker17,socolovsky19}).

\subsection{Structural evolution by star formation in the proto-cluster environment 
\label{subsec:evolution}}

In this subsection, we discuss the subsequent evolution of the total stellar mass 
and the stellar mass surface density at $r<1$~kpc ($\Sigma_1$) by continued star formation for the two sub-samples in USS1558.

\citet{barro17} suggested that 
star-forming galaxies in a normal star-formation phase 
and those in a compaction phase follow different evolutionary tracks 
on the log($\rm M_*$) -- log($\rm \Sigma_{1}$) diagram. 
When star-forming galaxies evolve in the inside-out manner, 
they likely move on this diagram along the relation 
of $\rm log\ \Sigma_{1} \propto 0.9\ log\ M_*$. 
When galaxies are in a compaction phase, 
in which bulge-like structures are rapidly built up at the center, 
their evolutionary track becomes steeper than that of the inside-out growth, 
namely, $\rm log\ \Sigma_{1} \propto 1.3\ log\ M_*$, to move the sequence of quiescent populations.

Assuming a constant SFR with time, 
we calculate how the total stellar mass and $\rm \Sigma_{1}$ changes  
using the total $\rm SFR_{H\alpha, corr}$ and $\Sigma_{\rm SFR}$ at $r<1$~kpc. 
We calculate $\rm \Delta log\ \Sigma_{1}/\Delta log\ M_*$ 
with $\Delta t =$ 500~Myr as an example case.  
As a result, we obtain $\rm \Delta log\ \Sigma_{1}/\Delta\ log\ M_* = 0.68$ 
for the high-mass sub-sample   
and $0.95$
for the low-mass sub-sample, respectively. 
The results do not depend on $\Delta t$. 
The low-mass sub-sample seems to increase the central mass surface density  
and total stellar mass along the evolutionary path of galaxies 
in a normal star-formation phase 
with $\rm \Delta log\ \Sigma_{1}/\Delta log\ M_* = 0.9-1.0$ 
\citep{tacchella16a,barro17}. 
As for the high-mass sub-sample, 
the central mass surface density seems to grow more slowly 
than the total stellar mass, 
and the values of $\rm \Delta log\ \Sigma_{1}/\Delta log\ M_*$ are 
close to the slope of the $\rm log\ M_*$--$\rm log\ \Sigma_{1}$ relation 
for quiescent galaxies 
\citep{fang13,barro17}.
Such a flatter evolutionary path is suggested in the simulations 
for galaxies in the inside-out quenching phase after the compaction event 
\citep{zolotov15,tacchella16a}. 
However,  
the sSFR profile of the high-mass sub-sample shows 
a weak decline toward the center rather than a strong suppression at the center
(the right panel of figure~\ref{fig:radialprofile}). 
Considering compact dust emission observed at the center 
of massive star-forming galaxies at high redshifts (e.g., \cite{barro16,tadaki17}), 
the shallower slope of the high-mass sub-sample may be explained 
by dust-obscured star formation which cannot be fully recovered in our method with 
UV and H$\alpha$.

The obtained results shown in subsection~\ref{subsec:image}, \ref{subsec:comparesize} 
and $\rm \Delta log\ \Sigma_{1}/\Delta log\ M_*$ values  
suggest that our samples in the dense proto-cluster core 
have an extended star-forming disk 
and evolve their structures in a secular way from inside to outside. 
We find no clear sign of the on-going compaction event 
in the stacked SFR or sSFR profiles, 
implying that gas-rich mergers or violent disk instabilities 
may be sub-dominant processes on average in the high-density region at $z\sim2.5$. 
%%%%
In Shimakawa et al.\ (\yearcite{shimakawa17}, \yearcite{shimakawa18}), 
they reported  
that star-forming galaxies in the dense groups in USS1558 have  
systematically higher sSFRs than those in the less dense regions  
and that such high star-forming activity may be supported by 
a large amount of {\sc Hi} gas residing in the proto-cluster core. 
%%% 
Our results about the spatial extent of the star-forming region within them  
suggest that their star-forming activity is enhanced across the entire disk 
rather than concentrated only at the central region \citep{nelson16b}. 
Their extended star-forming disks may be maintained by 
continuous, steady gas inflow from outside 
(e.g., Dav\'e et al. \yearcite{dave11_1}) keeping its angular momentum.

Our results also indicate that gas removal processes, 
such as ram pressure stripping, 
may not be acting on star-forming galaxies effectively in the proto-cluster core at $z=2.53$, in contrast to
local high-density environments 
where star-forming galaxies tend to show truncated star-forming disks 
(e.g., \cite{koopmann04a,schaefer18}). 
Because the proto-cluster is in vigorous assembly phase and the potential well is likely still immature as compared to the local clusters 
\citep{chiang17},  
interactions with hot gas in the cluster core may not be active yet.

Note that
we show only the average trends among our sample based on the stacking analyses. 
We might observe a centrally concentrated star-forming region 
for some of our targets 
if we look at the individual galaxies. 
Investigating the internal distribution of star-forming regions in the individual galaxies 
is necessary to evaluate the relative contributions among 
different physical processes and their environmental dependence. 
Moreover, 
considering a high fraction of low-mass ($\rm log(M_*/M_\odot) \lesssim 9$) galaxies 
with elevated SFRs in this proto-cluster field \citep{hayashi16}, 
such low-mass galaxies may hold key information on the early environmental dependence.
Investigation of internal structures in those low mass systems must therefore be very important, although
such studies require much deeper imaging data.

\section{Summary}

We conducted the AO-assisted $K'$-band and NB imaging observations 
with Subaru/IRCS+AO188 
for the star-forming galaxies in the dense proto-cluster core 
at $z=2.53$. 
By combining AO and the NB filter, 
we were able to spatially resolve the H$\alpha$-emitting regions within the galaxies. 
We obtained the images for 11 H$\alpha$ emitters 
with an angular resolution (FWHM) of 0.25~arcsec, 
corresponding to 2~kpc at $z=2.53$.

We conducted the stacking analyses by dividing the sample into two stellar mass bins, 
namely, the high-mass sub-sample with $\rm log(M_*/M_\odot) = 10.0-11.1$ 
and the low-mass sub-sample with $\rm log(M_*/M_\odot) = 9.5-10.0$. 
With the stacked images, 
we compared the spatial distribution of star-forming regions 
and the underlying stellar components. 
Our findings are the following: 

\begin{itemize}
    \item The sizes of the stellar components are estimated to be 
    $R_{1/2} = 3.34^{+0.16}_{-0.15}\ {\rm kpc}$ and $3.70^{+0.45}_{-0.39}\ {\rm kpc}$ 
    for the high-mass and low-mass sub-sample, respectively, 
    when assuming the radially dependent dust extinction for H$\alpha$. 
    The sizes of the star-forming regions traced by H$\alpha$ are estimated to be 
    $R_{1/2} = 4.13 \pm 0.11\ {\rm kpc}$ and $4.02 \pm 0.23 \ {\rm kpc}$ 
    for the high-mass and low-mass sub-sample.
    The high-mass sub-sample shows more extended star forming-region than the underlying stellar 
    component.

    \item Comparing our results for the proto-cluster
    with those for the field galaxies at $z=2-2.5$ on the stellar mass--$R_{1/2}$ diagram, 
    we found no clear environmental dependence at least for relatively massive galaxies. 
    They have more extended star-forming regions than stellar components irrespective of the environments. 
    Our low-mass sub-sample in the proto-cluster has slightly larger $R_{1/2}$ than the field counterparts, 
    but this may be in part related to systematically high sSFRs for the low-mass sub-sample in USS1558. 
    
    \item We investigated the growth of the total stellar mass and central mass surface density at $r<1$~kpc 
    by assuming a constant SFR during a given time period.
    The high-mass and low-mass sub-samples show $\rm \Delta log(\Sigma_{1})/\Delta log(M_*) \sim$ 0.68 and 0.95, respectively. 
    These values indicate that the two sub-samples likely grow their structures 
    in an inside-out manner rather than by a compaction event \citep{barro17}. 

\end{itemize}

Our results suggest that the structural growth of star-forming galaxies at $z=2-2.5$ 
is dominated by internal secular processes even in the dense proto-cluster core, 
and that galaxies at least with $\rm log(M_*/M_\odot) \ge 9.5$ 
are forming stars over their entire disks.

%% Future prospects??
Deep AO-assisted IFU observations to detect multiple emission lines 
will enable us to investigate radial profiles of [{\sc Nii}]/H$\alpha$ and H$\alpha$/H$\beta$ ratios 
for more accurate estimations of SFR distributions. 
Additionally, tracing the dust emission distribution by rest-frame infrared observations is also required 
to uncover the star-formation activity that is largely obscured in the rest-frame UV-optical regime. 
The Atacama Large Millimeter/submillimeter Array (ALMA) will allow us to map 
such dusty star-forming regions with the same (or higher) angular resolution as that of the AO-assisted observations 
(e.g., \cite{barro16,tadaki17}). 
Also, mapping molecular gas components within galaxies with ALMA 
will be crucial to investigate the presence of the compaction event more directly.

\begin{ack}
We thank the anonymous referee for careful reading and comments that 
improved the clarity of this paper. 
We would like to thank the Subaru telescope staff for supporting the observations. 
This work is based on observations made with the NASA/ESA Hubble Space Telescope, 
obtained at the Space Telescope Science Institute, which is operated by the Association of
Universities for Research in Astronomy, Inc., under NASA contract NAS 5-26555. 
These observations are associated with program GO-13291.
Data analyses were in part carried out on the open use data
analysis computer system at the Astronomy Data Center, ADC, of 
the National Astronomical Observatory of Japan (NAOJ). 
A part of this study is conducted with the Tool for OPerations on Catalogues And Tables (TOPCAT; \cite{topcat}).
TK acknowledges support by JSPS KAKENHI Grant Number JP18H03717.
\end{ack}

\section*{Appendix 1. Comparison of total fluxes between IRCS and MOIRCS}

\begin{figure}
\begin{center}
    \includegraphics[width=0.9\columnwidth]{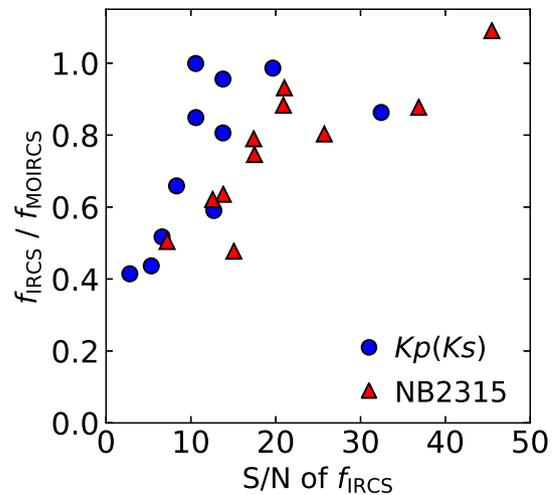}
    \end{center}
    \caption{Flux density ratio between IRCS ($K'$ and NB2315) and MOIRCS ($K_s$ and NB2315) as a function of 
    S/N of the IRCS total flux density for 11 H$\alpha$ emitters analyzed in this study.
    The total flux density ratios become smaller with decreasing S/N, 
    indicating that the extended components for the faint emitters tend to be missed in the IRCS+AO188 images.   
    }
    \label{fig:fluxloss}
\end{figure}

Figure~\ref{fig:fluxloss} shows a comparison of total flux densities 
between the IRCS $K'$-band/NB2315 images (as mentioned in subsection~\ref{subsec:stacking}) 
and the MOIRCS $K_s$-band/NB2315 images \citep{hayashi16}
as a function of S/N of the flux densities of the IRCS images. 
The total fluxes for the MOIRCS images represent the Korn fluxes measured with SExtractor \citep{sextractor}. 
The IRCS flux densities 
tend to be smaller than the MOIRCS flux densities at lower S/N, 
indicating that some fluxes are missed in our IRCS+AO188 imaging observations. 
This is probably caused by the lower throughput and higher thermal background 
of IRCS than MOIRCS. 
These factors lead to lower sensitivity of IRCS especially for diffuse components. 
As a result, 
the depths of the current data are insufficient 
to detect extended emission from the H$\alpha$ emitters.

The average value $f_{\rm IRCS}/f_{\rm MOIRCS}$ in the $K'_{(s)}$-band (NB2315) image 
is 84 (86)\% for the high-mass sub-sample and 
65 (68)\% for the low-mass sample. 
The missed fluxes are less than 20\% for most objects in the high-mass sub-sample. 
We consider that the missed fluxes do not strongly affect our results obtained from the stacking analyses 
for the high-mass sub-sample. 
As for the low-mass sub-sample, flux loss may lead to underestimation of size. 
However, when comparing the size of the stacked $K'$-band image  
with that measured from the stacked $H_{F160W}$-band image with GALFIT (subsection~\ref{subsec:comparesize}), 
the size in our measurement is rather larger than the GALFIT result. 
We unlikely underestimate the size due to the flux loss, and 
the systematic offset in our size measurement seems to affect the result dominantly.

\end{document}